\documentclass[12pt, A4]{article}
\thicklines                     
\begin{document}
\begin{titlepage}
\begin{centering}
\vspace{2cm}
{\Large\bf The energy-carrying velocity and rolling of tachyons of unstable
  D-branes}\\
\vspace{2cm}
Jin~Hyun Chung$^a$\footnote{E-mail address: {\tt
    jhchung@mail.chosun-c.ac.kr}}
and 
W.~S.~L'Yi$^b$\footnote{E-mail
address: {\tt wslyi@hep.chungbuk.ac.kr}}\vspace{0.5cm}

$^a${\em Department of Liberal Arts, Chosun College of Science \& Technology}\\
{\em 290 Seosuk Dong, Dong-gu Kwang-ju City, Korea,}\\\vspace{0.2cm}
$^b${\em Department of Physics, Chungbuk National University,}\\
{\em Cheongju, Chungbuk 361-763, Korea.}\\
\vspace{1cm}
\begin{abstract}
\noindent 
It is shown that the tachyons which originate from unstable
D-branes carry energy and momentum at the velocity $\beta = c^2/v,$ 
where $v$ is the phase velocity which is greater than $c.$ 
For an observer who moves with the velocity $\beta,$ tachyon is
observed to be moving from one of the ground states
of the tachyon potential to the potential hill.  It is found that
tachyon either passes over the hill or bounces back to the original
ground state. Another possible solution is the case
which is margial to these, that is tachyon reaches to
the top of the potential hill and stays there forever.
\end{abstract}
\end{centering}
\vspace{.5cm}
\vfill

Nov.13, 2003\\
\end{titlepage}

\setcounter{footnote}{0}
\section{Introduction}
Tachyons, which abound in non-BPS D-branes of type
IIA/IIB superstring theories or in bosonic strings, can be studied by using
relatively simple effective action\cite{effective_action}. The tachyon
potential $V(T),$ which is an even function of the tachyon field $T,$
has a run-away form: it is maximum at $T=0,$ and minimum at
$T\to\pm\infty.$  This means that if the initial string state is the tachyonic state
around the unstable false vacuum $T\approx 0,$ all the energy, because
of the fact that the full string states contain various field modes, 
will eventually be distributed
among various bulk modes.

This process of rolling of tachyons and tachyon condensations are discussed
by various authors\cite{rolling}.  Recently it is pointed out that
fluctuations around a static kink solution of D$_p$\,-brane produces the
Dirac-Born-Infelt action of a BPS D$_{p-1}$\,-brane\cite{DBI_kink}. It
is also found\cite{Brax} that the
only  static finite energy solution of tachyon field which depends
only on  one of the spatial coordinate $x$ is $T(x) = Ax,$ where
$A\to\infty.$

In this paper, we investigate the traveling wave solutions of the tachyon field
which propagate along $x.$ It is found that the phase velocity $v$ of
tachyon wave may have any value except the speed of light $c.$ As
usual, the phase velocity is quite different from the mechanical
velocity\footnote{The usual terminology for the velocity at which
waves carry the energy-momentum is the group veolcity.  But in this
tachyonic case whose highly non-linear equation of motion does not
admit the simple rule of linear superposition, it is more adequate to use the term 
mechanical velocity instead of group velocity.} 
at which the energy-momentum is carried. If $v<c,$ thay are
the same when one assumes that the wave amplitude $A$ becomes very
large.   On the other hand, if $v>c,$ the mechanical velocity is $\beta = c^2/v.$
Furthermore, the invariant mass $m=\sqrt{E^2-p^2c^2}/c^2$ is always
real. This important fact is pointed out in\cite{SJR} where
boundary state idea is used under different perspective.
It means that tachyonic states of unstable D-branes can be interpreted
as usual matter states moving at the mechanical velocity $\beta.$ This
energy distribution in space may be related to the final bulk states
of strings.

For an observer who moves with the velocity $\beta,$ tachyon is
observed to be rolling from one of the ground states
of the tachyon potential to the potential hill.  
It is found that there are three different types of motion.
The first is that tachyon passes over the hill and rolls to another
ground state.  The second one is the case that it can not cross
over the hill and bounces back to the original ground state. 
The last possibility is margial to these, and tachyon reaches to
the top of the potential hill and stays there forever.

In section 2, a propagating wave solution for the tachyon is obtained.
It has either oscillating form or kink form.  The oscillating solution
is studied in section 3, and the kink solution is investigated in
section 4. Physical interpretation of the solution is given if section
5. The conclution is given in section 6.

\section{Propagating wave solution for the tachyon field}

Consider the following effective action of the tachyon field $T$
on a non-BPS D$_p$\,-brane,
\begin{equation}
S= - \int d^{p+1} x \;  V(T) \sqrt{ -\det(\eta_{\mu\nu} 
	+ \partial_\mu T \partial_\nu T)}.
\end{equation}
Here, $\eta_{\mu\nu} = {\rm diag}
(-1, 1,...,1),$ and $\hbar=c=\alpha'=1.$ For the potential of the
tachyon field, we assume the following form\cite{{potential_form}, {Kutasov1}}
\begin{equation}
V(T) = {\tau_p \over \cosh ( {T \over T_0})},
\end{equation}
where $\tau_p$ is the brane tension, and $T_0$ is a constant which is
$\sqrt{2}$ for non-BPS D-branes in superstrings, and 2 for bosonic strings.

To solve the Lagrangian equation of motion, we introduce a new variable ${\cal T}$ defined 
by\cite{{Kutasov1}, {new_variable_T}}
\begin{equation}
{\cal T} = T_0 \sinh {T \over T_0}.\label{def_cal_T}
\end{equation}
The Lagrangian in terms of ${\cal T},$ which is given by
\begin{equation}
{\cal L} = - {\tau_p\over 1 + {\displaystyle{\cal T}^2 \over \displaystyle T_0^2}} 
            \sqrt{ 1 + {{\cal T}^2 \over T_0^2}
	+ \partial_\mu{\cal T} \partial^\mu{\cal T}}, 
\end{equation}
does not contain any transcendental function.
The corresponding equation of motion is given by the following
polynomial form of ${\cal T},$ $\partial_\mu{\cal T}$ and $\partial_\mu\partial_\nu{\cal T}:$
\begin{equation}
( 1 + {{\cal T}^2 \over T_0^2}) (\partial_\mu\partial^\mu {\cal T} + { {\cal T} \over T_0^2})
	+ \partial_\mu{\cal T} \partial^\mu{\cal T}\,\partial_\nu\partial^\nu {\cal T} 
	- \partial_\mu{\cal T} \partial_\nu{\cal
	T}\,\partial^\mu\partial^\nu {\cal T} = 0.
\label{em}
\end{equation}
We observe the following facts that in each term of this equation, when
expanded, $\partial_\mu$ appears even times while ${\cal T}$ odd
times.  This means that for a given solution, one may obtain
another solution by changing either the sign of a paticular $x^\mu$ or/and ${\cal T}.$

It is known\cite{DBI_kink} that for the extreme {\sl static} kink solution such as
$T\to\pm\infty$ depending on the sign of $x^p,$ the resulting
configuration is a stable D$_{p-1}$\,-brane at
$x=0.$  In this paper we assume that ${\cal T},$ or equivalently
$T,$ depends on both $x=x^p$ and $t.$  In this case, (\ref{em}) can be written as
\begin{equation}
( 1 + {{\cal T}^2 \over T_0^2}) (\partial_x^2 - \partial_t^2 + { 1\over T_0^2}){\cal T}
   + 2\partial_x{\cal T} \partial_t{\cal T}\,\partial_x\partial_t {\cal T} 
   - (\partial_x{\cal T})^2\, \partial_t^2{\cal T}
   - (\partial_t{\cal T})^2\, \partial_x^2{\cal T}
 = 0.
\end{equation}
It is easy to show that it has the following traveling wave solution,
\begin{equation}
{\cal T} = {T_0\over 2}\left\{A{\rm e}^{ik (x-vt)} + B{\rm e}^{-ik
  (x-vt)}\right\},  \label{sol_osci_1}
\end{equation}
where $A,$ $B$ and $v$ are all integration constants, and $k$ is given by 
\begin{equation}
k = {1\over \sqrt{1- v^2} } {1 \over T_0}.
\end{equation}
If $|v| <1,$ $k$ is real and the solution is oscillatory. This is
discussed in section 3. On the other hand, if $|v| > 1,$ one may replace $ik$ by
$\kappa$ to obtain a kink solution.  It is discussed in section 4.

\section{Traveling sinusoidal solution}

Consider the case $|v| < 1.$ For the convenience of computation we
rewrite (\ref{sol_osci_1})  in the following form
\begin{equation}
{\cal T} = T_0 A \cos k(x-vt).\label{sol_cos}
\end{equation}
To understand the physical meaning of the phase velocity $v,$ we
calculate the energy and momentum densities of the field
configuration. The relevant energy-momentum tensor $T^\mu{}_\nu$ is 
\begin{equation}
T^\mu{}_\nu= -{\partial {\cal L} \over \partial \partial_\mu
  T}\,\partial_\nu T + \delta^\mu{}_\nu {\cal L},
\end{equation}
where the Lagrangian expressed in terms of $T(x,t)$ is given by 
\begin{equation}
{\cal L} = - V(T)\sqrt{1 + (\partial_x T)^2 - (\partial_0 T)^2}.
\end{equation}

From this one can read off the following energy density
$\rho=T^{00}$ and momentum density $P=T^{p0},$
\begin{eqnarray}
\rho &=&  {V(T) \over \sqrt{ 1+ (\partial_x T)^2 - (\partial_0 T)^2} } 
          \left\{ 1 + (\partial_x T)^2 \right\}, \label{gen_rho} \\
P &=& {V(T) \over \sqrt{ 1+ (\partial_x T)^2 - (\partial_0 T)^2} }
          \;\partial_x T \partial_0 T. \label{gen_P}
\end{eqnarray}
After non-trivial but straight forward calculations it can be shown
that 
\begin{equation}
 {V(T) \over \sqrt{ 1+ (\partial_x T)^2 - (\partial_0 T)^2} }  =
 {\tau_p \over \sqrt{1 + A^2}}.
\end{equation}
On the other hand, by using the defining equation (\ref{def_cal_T}) and the solution
(\ref{sol_cos}), one may prove that
\begin{eqnarray}
(\partial_x T)^2 &=&{(\partial_x {\cal T})^2 \over 1 + {\displaystyle
            {\cal T}^2 \over  \displaystyle T_0^2}} \\ 
            &=& k^2 T_0^2  { A^2 \sin^2 k(x-vt) \over 1 + A^2 \cos^2 k(x-vt)}.
\end{eqnarray}
That is, $\rho$ and $P,$ for this traveling sinusoidal solution, are
\begin{eqnarray}
\rho &=& {\tau_p \over \sqrt{1 + A^2} }
\left\{ 1 + {1\over 1 - v^2}  { A^2 \sin^2 k(x-vt) \over 
1 + A^2 \cos^2 k(x-vt)}\right\},\\
P &=&{\tau_p \over \sqrt{1 + A^2} }
{v\over 1 - v^2}  { A^2 \sin^2 k(x-vt) \over 1 + A^2 \cos^2 k(x-vt)}.
\end{eqnarray}
When $A=0$, one has the trivial unstable solution, $T=0.$ In this
case, $\rho = \tau_p$ and $P=0$ as expected.

For general $A,$ the energy density is oscillatory at the string length scale $T_0.$  
The average behaviour of it can be obtained by using the relation,
\begin{equation}
{1\over 2\pi} \int^{2\pi}_0 {a\sin^2\theta \over 1 + a \cos^2\theta}d\theta = \sqrt{ 1+a} -1.
\end{equation}
That is, the spatial average of $\rho$ is
\begin{equation}
\bar{ \rho} = {\tau_p \over 1 -v^2 } \left( 1 -{ v^2 \over \sqrt{1 + A^2}}\right).
\end{equation}
For large $A,$ it becomes
\begin{equation}
\bar{\rho} \simeq {\tau_p \over 1 - v^2}.
\end{equation}
It can be interpleted in the following way. The energy stored in
length $L$ is  $E_L = \bar{\rho} L.$ Here, one should be careful that it is not $L$ but $L_0 =
\gamma L$ which is Lorentz invariant, where $\gamma = 1/\sqrt{1 -v^2}.$  
It means that it is more appropriate to express $E_L$ in terms of $L_0,$  
\begin{equation}
E_L = \gamma \tau_p  L_0. \label{E_L}
\end{equation}
This shows that the (rest) mass density is $\rho_m = \tau_p. $

Similarly, the average momentum density is
\begin{equation}
\bar{ P} = {\tau_p v \over 1 - v^2} \left( 1 - {1 \over { \sqrt{1+A^2}}}\right).
\end{equation}
This means that, for large $A,$ the momentum in length $L$ is
\begin{eqnarray}
P_L &=& \bar{ P} L \\
    &=& \gamma \tau_p v L_0. \label{P_L}
\end{eqnarray}
Combining this with (\ref{E_L}) one gets $P_L = v E_L.$ It shows that
$v$ is in fact the velocity at which the energy is propagated.
Similarly we have $E_L^2 - P_L^2 = m_L^2,$ where $m_L = \tau_p L_0$ is
the mass stored in the invariant length $L_0.$

\section{Traveling kink solution}

Consider the case $|v|>1.$  In this case, by replacing $ik$ of (\ref{sol_osci_1})
with $\kappa$ we obtain the following solution,
\begin{equation}
{\cal T} = {T_0 \over 2}\left\{A {\rm e}^{\kappa(x-vt)} + B {\rm e}^{-\kappa(x-vt)}\right\}.
\label{sol_kink}
\end{equation}
Here, 
\begin{equation}
\kappa = {1\over \sqrt{v^2 -1 }}{1 \over T_0}.
\end{equation}
The corresponding energy and momentum densities which can be determinded from
(\ref{gen_rho}) and (\ref{gen_P}) are 
\begin{eqnarray}
\rho &=& {\tau_p\over \sqrt{1 + AB }}\left\{1+(\partial_x T)^2\right\},\\
P    &=& {\tau_p  \over \sqrt{1 + AB }}(\partial_x T)^2\; v.
\end{eqnarray}

By using the same technique employed in section 3, one may prove that
\begin{equation}
(\partial_x T)^2 = \kappa^2 T_0^2 {A^2 {\rm e}^{2\kappa(x-vt)}
                              + B^2{\rm e}^{-2\kappa(x-vt)}   -2AB  
                               \over 
                               A^2{\rm e}^{2\kappa(x-vt)} + B^2 {\rm e}^{-2\kappa(x-vt)}
                               + 2AB+4}.
\end{equation}
That is, 
\begin{eqnarray}
\rho &=& {\tau_p \over \sqrt{1 + AB}}
\left\{ 1 + \kappa^2 T_0^2 
{ A^2 {\rm e}^{2\kappa (x-vt)} + B^2 {\rm e}^{-2\kappa (x-vt)} -2AB  
  \over 
  A^2 {\rm e}^{2\kappa (x-vt)} + B^2 {\rm e}^{-2\kappa (x-vt)} + 2AB+4} \right\}, 
\label{rho_kink}\\
P &=& {\tau_p \over \sqrt{1 + AB}}\;
 \kappa^2 T_0^2 
{ A^2 {\rm e}^{2\kappa (x-vt)} + B^2 {\rm e}^{-2\kappa (x-vt)} -2AB  
  \over 
  A^2 {\rm e}^{2\kappa (x-vt)} + B^2 {\rm e}^{-2\kappa (x-vt)} + 2AB+4}  v. \label{P_kink}
\end{eqnarray}
To have real $\rho$ and $P,$ it is required that $AB > -1.$
We consider $AB>0,$ and $0 > AB >-1,$ $AB=0$ cases seperately.

\vspace{.25in}

 Case 1) $AB > 0.$

In this case, one may assume that $A=B.$  It is always possible
by properly choosing the origin of $x.$
The solution has the following form
\begin{equation}
{\cal T} = T_0 A \cosh\kappa(x-vt).
\end{equation}
The energy density $\rho$ which can be read off from (\ref{rho_kink}) is
\begin{equation}
\rho ={\tau_p \over \sqrt{1 + A^2} }
\left\{ 1 + {1 \over v^2 -1} { A^2 \sinh^2\kappa(x-vt) \over A^2 \cosh^2\kappa(x-vt) + 1} \right\}.
\end{equation}
This $\rho$ is almost constant except at the region of a small dimple of size $T_0.$

By using the relation
\begin{equation}
\int {a\sinh^2 x \over 1 + a\cosh^2 x} dx 
= x - \sqrt{1 + a} \tanh^{-1}\left\{{\tanh x \over \sqrt{1 + a}}\right\} + {\rm const.},
\end{equation}
one can show that the stored  energy in the region $-{L\over 2} < x <
{L\over 2}$ of length $L$ is the following,
\begin{equation}
E_L = {v^2\over v^2 - 1} {\tau_p \over\sqrt{ 1 + A^2}} L 
- {2\tau_p \over v^2 - 1} \tanh^{-1}\left( {\tanh{L\over 2} \over \sqrt{1+A^2}}\right).
\end{equation}
That is, the energy density $\bar{\rho} = \lim_{L\to\infty} E_L/L$ effectively becomes
\begin{equation}
\bar{\rho} = {\tau_p \over\sqrt{ 1 + A^2}} {1\over 1-\beta^2 } ,
\end{equation}
where $\beta = 1/v.$

Similarly, we have
\begin{equation}
\bar{P} = {\tau_p \over\sqrt{ 1 + A^2}} {\beta\over 1-\beta^2 } .
\end{equation}
It is interesting that the momentum density $\bar{P}$ vanishes if
$\beta \to 0,$ or equivalently if $v\to\infty.$ This gives us a hint
that $v$ is not the mechanical velocity.  In fact the
relation $\bar{P} =\beta\bar{\rho}$ shows that it is not $v$ but $\beta$ which carries energy.
This means that tachyon can be interpreted as a usual matter which
moves at the mechanical velocity $\beta.$ Similar analysis done in section 3 may be
employed here to show that the mass density is
\begin{equation}
\rho_m = {\tau_p \over \sqrt{1+A^2}}.
\end{equation}

\vspace{.25in}
Case 2) $0> AB >-1.$

By a suitable choice of the origin of $x$ one may assume that $B = -A,$ where $0<A^2<1.$
Then, the solution has the following form,
\begin{equation}
{\cal T} = T_0 A \sinh\kappa(x-vt).
\end{equation}
The corresponding energy density $\rho$ is
\begin{equation}
\rho ={\tau_p \over \sqrt{1 - A^2} }
\left\{ 1 + {1 \over v^2 -1} { A^2 \cosh^2\kappa(x-vt) \over A^2 \sinh^2\kappa(x-vt) + 1} \right\}
\end{equation}
As the case 1), $\rho$ has a small dimple of size $T_0.$
By using the relation
\begin{equation}
\int {a\cosh^2 x \over 1 + a\sinh^2 x} dx 
= x - \sqrt{1 - a} \tanh^{-1}\left\{{\sqrt{1-a}\tanh x }\right\} + {\rm const.},
\end{equation}
it can be shown that the total energy and momentum stored in the
region $-{L\over 2} < x < {L\over 2}$ of length $L$ are  
\begin{eqnarray}
E_L &=& {1\over 1- \beta^2} {\tau_p \over\sqrt{ 1 - A^2}} L 
- {2\tau_p \beta^2\over 1 - \beta^2 } \tanh^{-1}\left(\sqrt{1 - A^2}\tanh
{L\over 2}\right),\\
P_L &=& {\beta\over 1- \beta^2} {\tau_p \over\sqrt{ 1 - A^2}} L 
- {2\tau_p \beta\over 1 - \beta^2 } \tanh^{-1}\left(\sqrt{1 - A^2}\tanh
{L\over 2}\right).
\end{eqnarray}
That is, the energy and momentum densities, when $L\to\infty,$ effectively become
\begin{eqnarray}
\bar{\rho} &=& {1\over 1 -\beta^2} {\tau_p \over\sqrt{ 1 - A^2}},\\
\bar{P} &=& {\beta\over 1 -\beta^2} {\tau_p \over\sqrt{ 1 - A^2}}.
\end{eqnarray}
These show that the tachyon can in fact be inperpreted as a usual matter of mass density 
\begin{equation}
\rho_m = {\tau_p \over \sqrt{1-A^2}}
\end{equation}
moving at the velocity $\beta = 1/v.$

\vspace{.25in}
Case 3) $AB=0.$

We assume that $B=0.$  The other case, that is $A=0,$ can be obtained
from the results of this case by replacing $\kappa \to -\kappa.$ The
corresponding solution is given by
\begin{equation}
{\cal T} = A {\rm e}^{\kappa (x -vt)}.
\end{equation}

The energy and momentum densities are
\begin{eqnarray}
\rho &=& \tau_p\left\{  1 + {1 \over v^2-1} {A^2 {\rm e}^{2\kappa(x-vt)} \over A^2 {\rm e}^{2\kappa(x-vt)} 
       + 4}\right\}, \\
P &=& \tau_p \left\{ {1 \over v^2-1} {A^2 {\rm e}^{2\kappa(x-vt)} \over A^2 {\rm e}^{2\kappa(x-vt)} 
       + 4}\right\}v. \\
\end{eqnarray}
These mean that the energy and momentum densities under the $A \to \infty$ limits become
\begin{eqnarray}
\rho &=& {\tau_p \over 1 - \beta^2},\\
P &=& {\tau_p \over 1 - \beta^2}\beta.
\end{eqnarray}
This means that it also describes the usual matter of the mass density
$\rho_m = \tau_p$ moving at the velocity 
$\beta = 1/v.$

\section{Interpretation of the solutions}
Consider the traveling sinusoidal solution (\ref{sol_cos}) first.
For an observer with coordinates $(x',t')$ who moves at the velocity
$v$ with respect to the
frame $(x,t),$ the solution can be written as
\begin{equation}
{\cal T} =T_0 A\cos{x' \over T_0}.
\end{equation}
This represents an infinite lattice of brain-antibrane system\cite{lattice}.

For the traveling kink solution, consider (\ref{sol_kink}).
For an observer with coordinates $(x',t')$ who moves with velocity $\beta,$ phase
$\kappa(x-vt)$ can be written as $-{t' \over T_0},$
where we used 
\begin{equation}
{x-vt \over \sqrt{v^2-1}} = -t'.
\end{equation}
This means that for kink solution it can be written as
\begin{equation}
{\cal T} = \left\{ 
   \begin{array}{ll}
T_0 A\cosh(-{t'\over T_0}) & \mbox{for Case 1,}\\
T_0 A\sinh(-{t'\over T_0}) & \mbox{for Case 2,}\\
T_0 A{\rm e}^{-{t'\over T_0}} & \mbox{for Case 3.}
   \end{array}
\right.
\end{equation}      

It is clear that Case~2 represents motion of tachyon which moves
from one ground state of the the potential at $t'\to-\infty$ to the 
other ground state at $t'\to\infty,$ thus crossing over the potential hill.
For the Case~1, tachyon does not cross over the potential hill, and
begins to bounce back to the original vacuum at $t'=0.$
The Case~3 is marginal to these, and tachyon approaches to the
maximum of the potential as $t'\to\infty.$

\section{Conclusion}
The traveling wave solution of the Sen's tachyon action, either in the
oscillating form or kink form, represents usual matter that moves at
the velocity $\beta $ which is less than $c.$ If the phase velocity $v$ is less than $c,$
the tachyon field oscillates in space.  In this case, $\beta = v$ and the mass
density is $\tau_p.$

If, on the other hand, the phase velocity of the tachyon field $v$ is
greater than the speed of light $c,$ tachyon field forms a kink.  In
this case, the velocity at which energy is carried is $\beta = c^2/v.$
This corresponds to the rolling of tachyons.
The corresponding mass density is $\tau_p/\sqrt{1+ A^2},$
$\tau_p/\sqrt{1 - A^2}$ or
$\tau_p$ depending on the precise form of the solution. This
energy-momentum may be distributed among various modes of strings.

\vspace{.25in}
\noindent{\bf \Large Acknowledgement}

\noindent
The author would like to thank Paul Frampton
for his interests and critical comments on this paper, and for his hospitality during
the author's stay in UNC, Chapel Hill.

\end{document}